\documentclass[prb,showpacs,twocolumn,aps]{revtex4}
\usepackage{graphicx}
\newcommand{\be}{\begin{equation}}
\newcommand{\ee}{\end{equation}}
\newcommand{\bea}{\begin{eqnarray}}
\newcommand{\eea}{\end{eqnarray}}
\newcommand{\bwt}{\begin{widetext}}
\newcommand{\ewt}{\end{widetext}}

\def\H{{\cal H}}
\newcommand{\D}{{\cal D}}

\newcommand{\etal}{{\it et al.} }
\newcommand{\pg}{{pseudogap} }
\newcommand{\bz}{{Brillouin zone}}

\begin{document}
\title{Phase fluctuations of $s$-wave superconductors on a lattice}
\author{Wonkee Kim$^{1}$ and J. P. Carbotte$^{2}$}
\affiliation{
$^{1}$Department of Physics, University of Alberta, Edmonton, Alberta,
Canada, T6G~2J1\\
$^{2}$Department of Physics and Astronomy,
McMaster University, Hamilton,
Ontario, Canada, L8S~4M1}
\begin{abstract}
Based on an attractive $U$ Hubbard model on a lattice with up to
second neighbor hopping we derive an effective Hamiltonian for phase 
fluctuations. The superconducting gap is assumed to have $s$-wave symmetry.
The effective Hamiltonian we finally arrive at is of the extended XY
type. While it correctly reduces to a simple XY in the continuum limit, in 
the general case, it contains higher neighbor interaction in spin
space. An important feature of our Hamiltonian is that it gives a much
larger fluctuation region between the Berezinskii-Kosterlitz-Thouless
transition temperature identified with $T_{c}$ for superconducting
and the mean field transition temperature identified with the pseudogap 
temperature.
\end{abstract}
\maketitle

\section{introduction}

The origin of the \pg phenomena\cite{timusk} in high-$T_{c}$ superconductors
is one of the most challenging questions in the theory of superconductivity
but there is no consensus on the correct theoretical approach to 
be taken to describe such phenomena.
What is generally accepted is that the \pg is a manifestation of
strong correlation effects which become progressively more important
as the doping is reduced into the underdoped regime and the
Hubbard Mott insulating state is approached.
Roughly speaking, theories of the \pg state can be divided into two classes.

One is based on the idea of a precursor state to superconductivity.
In this scenario
the Cooper pairs are pre-formed below the \pg temperature $(T^{*})$
assumed larger than the superconducting transition
temperature $(T_{c})$ but there is no phase coherence
in the temperature range between $T_{c}$ and $T^{*}$.\cite{emery,carlson}
The phase fluctuations destroy phase coherence and
consequently
the superfluid stiffness is zero
in this temperature regime. Therefore, this picture envisions that
the two gaps ( superconducting and \pg) have the same origin.
The precursor state idea is based on the results of 
angle resolved photoemission spectroscopy (ARPES) experiments,\cite{ding}
which show
a smooth evolution of the pseudogap into the superconducting gap
as temperature is reduced. 
This is also consistent with tunneling results\cite{renner} of Renner
\etal who found a smooth evolution of tunneling characteristics as a function
of temperature showing the gap filling but not closing as one goes through
the superconducting transition temperature. On the other hand data by Krasnov
\etal\cite{krasnov} on intrinsic mesa junctions show a separate
superconducting peak and a \pg hump. These data do not favor a pre-formed
pair interpretation although superconducting and \pg both appear pinned to the Fermi energy.
A different precursor scheme\cite{levin} includes 
in the calculations finite momentum pairs
which exist at any finite temperature below $T^{*}$
but do not contribute to the superfluid density made up of zero momentum
Cooper pairs. 
Above $T_{c}$ there remain no zero
momentum pairs and therefore no superfluid density. This approach
constitutes a natural extension of BCS theory to the strong coupling
regime and has many features that agree with experiments on the cuprates.
For example \pg effects become more prominent as the doping is reduced into
the underdoped regime of the phase diagram.

A second class of theories
assumes that  there is a second order parameter which competes with
the superconducting gap and that the microscopic origin of the \pg 
associated with this second order parameter is
different from that of the superconducting gap. The competition between the
two gaps manifests itself most importantly in the underdoped 
regime and so in this doping
regime the two gaps co-exist. As the doping is increased towards optimum and 
into the overdoped regime the second order parameter is weakened and could even 
vanish in the overdoped case. This feature is quite different from the
precursor scenario.
A recent theoretical approach\cite{ddw}
based on this line of thought has assumed
that the \pg has a $d$-wave charge density wave order, the so-called
$d_{x^{2}-y^{2}}$-density wave (DDW).
An experimentally observed phase diagram\cite{tallon} for the high-$T_{c}$
cuprates is consistent with this competing gap picture. However, a more recent
theoretical work\cite{kim1} implies that the DDW model may not be 
consistent with
tunneling experiments.
Transport properties\cite{kim2} of the DDW state may also be used to
test this picture.

In this paper we deal only with the phase fluctuation picture.
CuO$_{2}$ planes play a crucial role in high-$T_{c}$
superconductivity and interlayer
coupling is weak. Here it is assumed that
the two-dimensional nature of the cuprates is an important feature of 
these compounds.
For simplicity, however, we will consider only $s$-wave pairing on a lattice
so that strictly speaking our work will not apply to high-$T_{c}$ cuprates
which are known to have $d$-wave symmetry. In this sense the present work
is a first essential step in a realistic treatment of the cuprates.

Noting that
spontaneous breaking of continuous symmetry is impossible
in 2D,\cite{mermin}
one may conceive\cite{practically} that the superconducting transition,
at least in underdoped cuprates, is not of the BCS kind
but is rather related to the
Berezinskii-Kosterlitz-Thouless (BKT) transition,\cite{kt} which is
found in two-dimensional systems.\cite{loktev,kim3,abrikosov1}
Of course, in reality, the coupling between CuO$_{2}$ planes
must become important right at $T_{c}$ since the transition must be
three-dimensional with true off-diagonal long range order.

The two-dimensional XY model\cite{jose}
on a square
lattice
is an example of the BKT transition. 
This model describes a system of spins ${\bf S}_{i}$ constrained to rotate
on the lattice, where $i$ indicates
the site. The Hamiltonian of the model can be 
written in the form
$-J\sum_{<ij>}{\bf S}_{i}\cdot {\bf S}_{j}$, where $J>0$ and
$<ij>$ means a sum over nearest neighbor pairs.
It has been proposed\cite{kt} 
that the BKT transition is associated with unbound
vortex-antivortex pairs. The mean-square distance 
$\langle r^{2}\rangle$
between a vortex
and an antivortex is
\be
\langle r^{2}\rangle=
{\pi J/T-1\over \pi J/T-2}\;.
\ee
As temperature approaches the BKT transition temperature $T_{BKT}=\pi J/2$,
$\langle r^{2}\rangle$ diverges and the pairs become unbound.

In a theoretical approach to the precursor scenario, 
Cooper pairs are
formed at the mean-field temperature $(T_{MF})$, which is identified
as $T^{*}$, 
The phase fluctuations
destroy the long range order
until the phases are locked in at the BKT
transition temperature $T_{BKT}$, which is $T_{c}$.
Consequently, the effective Hamiltonian for phase fluctuations
is mapped into a XY model
Hamiltonian.
In order to consider phase fluctuations of the superconducting order parameter,
one has to go beyond mean-field BCS theory. This can be implemented by
using the well-established path-integral formalism for 
fermionic fields.\cite{negele}
The effective phase Hamiltonian for
an $s$-wave superconductor can be described by $H_{XY}=J_{XY}
\int d{\bf r}(\nabla\theta)^{2}$, where $\theta$ represents
phase fluctuations. To obtain a lattice version of the XY Hamiltonian
one might discretize $H_{XY}$. However, it has been shown\cite{kim3} that
discretization does not pick up all the important features of the phase
Hamiltonian on a lattice, which will be derived and studied in detail
in this paper.

The paper is organized as follows:
In Section II, we introduce the
path integral approach and derive an effective action for phase
fluctuations. We apply the Hubbard-Stratonovich transformation
to integrate out the fermionic degrees of freedom. 
We also consider a case where the phase is treated in the continuum limit while
the fermions still remain
on a lattice. This serves as a bridge to the continuum
limit.
We show in section III that the
effective Hamiltonian is not of the usual XY type.
There are extra
terms which do not belong to the usual XY Hamiltonian.
It is shown how the new more complex Hamiltonian reduces to
the XY model in
the continuum limit. 
In this section we also explain how a periodic boundary condition applied to
the lattice leads directly to
a simple expression for the effective
Hamiltonian for phase fluctuations. 
In Section IV, we show that the effective
Hamiltonian turns out to be of an extended XY type and investigate
the effects of the extra terms. When we include next-nearest neighbor hopping 
in the electronic
energy dispersion (Section V), we find that the extended nature of the Hamiltonian
is preserved and the correction to the usual XY model becomes even more
essential.
We also explain a qualitative relation between the physics of the 
BCS-Bose Einstein (BE) crossover\cite{andrenacci}
and the BKT transition temperature.
Section VI contains discussions and conclusions.

\section{formalism}

We begin with a 2D attractive Hubbard model
on a square lattice in its simplest form conceivable, namely
\be
\H=-t\sum_{<ij>\sigma}C^{+}_{i\sigma}C_{j\sigma}
-\mu\sum_{i\sigma}n_{i\sigma}
-U\sum_{i}n_{i\uparrow}n_{i\downarrow}\;,
\ee
where $C_{i\sigma}$ is a fermion field with a spin $\sigma$,
$\mu$ is a chemical potential, $U(>0)$ is the pairing
interaction, $n_{i\sigma}=C^{+}_{i\sigma}C_{i\sigma}$, and
$t$ describes nearest neighbor hopping.
To start we consider only the nearest neighbor hopping for algebraic simplicity;
however, later we will  
include the next nearest neighbor hopping.
The symbol $<ij>$ means a sum carried out over nearest neighbor pairs
as mentioned earlier. 
When we need a double summation over $i$ and $j$, the usual notation
$\sum_{i,j}$ will be used. 
We will set the lattice constant $a=1$ and also use units such that
$\hbar=k_{B}=1$. 
The partition function $(Z)$ of this model in the path-integral formalism
is written as
\be
Z=\int\D C^{+}\D C \exp[-S]\;,
\ee
where the action $S=\int d\tau \{\sum_{i\sigma}C^{+}_{i\sigma}
\partial_{\tau}C_{i\sigma}+\H\}$ and the range for the integral
over the imaginary time
$\tau$ in the action is from $0$ to $1/T$. 

Defining spinors $\Psi_{i}$
for the fields $C_{i\uparrow}$ and $C^{+}_{i\downarrow}$ such that
$\Psi^{+}_{i}=(C^{+}_{i\uparrow}\; C_{i\downarrow})$, the action $S$
becomes
\bwt
\be
S=\int d\tau\Bigg\{
\sum_{i}\Psi^{+}_{i}\left[
{\hat\tau}_{0}\partial_{\tau}-{\hat\tau}_{3}\mu\right]\Psi_{i}
-t\sum_{<ij>}\Psi^{+}_{i}{\hat\tau}_{3}\Psi_{j}
-U\sum_{i}\Psi^{+}_{i}{\hat\tau}_{+}\Psi_{i}\Psi^{+}_{i}{\hat\tau}_{-}\Psi_{i}
\Bigg\}\;,
\ee
\ewt
where ${\hat\tau}_{\alpha}$ $(\alpha=0,1,2,3)$ are Pauli matrices and
${\hat\tau}_{\pm}=({\hat\tau}_{1}\pm i{\hat\tau}_{2})/2$. 
The standard manipulations to derive the effective action for phase fluctuations
includes: i) the Hubbard-Stratonovich transformation\cite{negele} 
to deal with the 
four-fermion-field term and ii) a gauge transformation to avoid violating
the Mermin-Wagner theorem.\cite{mermin}
In order to use the Hubbard-Stratonovich transformation, one needs to
introduce a complex auxiliary field
$\phi_{i}$ and
the partition function $Z$ becomes a function of $\Psi$ and $\phi$:
\bwt
\be
Z=\int \D\Psi^{+}\D\Psi\D\phi^{*}\D\phi \exp[-S]\;,
\ee
where
\be
S=\int d\tau\Bigg\{
{1\over U}\sum_{i}|\phi_{i}|^{2}+
\sum_{i}\Psi^{+}_{i}\left[
{\hat\tau}_{0}\partial_{\tau}-{\hat\tau}_{3}\mu\right]\Psi_{i}
-t\sum_{<ij>}\Psi^{+}_{i}{\hat\tau}_{3}\Psi_{j}
-\sum_{i}\left[\phi_{i}\Psi^{+}_{i}{\hat\tau}_{+}\Psi_{i}+h.c\right]
\Bigg\}\;.
\ee
Note that if we require the equation for $\phi^{*}_{i}$; namely
$\delta S[\phi,\phi^{*}]/\delta\phi^{*}_{i}=0$, the solution of this
equation is
$\phi_{i}=U C_{i\downarrow}C_{i\uparrow}$. Consequently, 
$\phi_{i}$ can be 
identified as the order parameter.

Parameterizing the auxiliary field $\phi_{j}=\Delta_{j}e^{i\theta_{j}}$
and making a gauge transformation
$\Psi_{j}=\exp[i{\hat\tau}_{3}\theta_{j}/2]\chi_{i}$,
where $\chi_{i}$ is a spinor for neutral fermions, one can
show that
\bea
S&=&\int d\tau\Bigg\{
{1\over U}\sum_{i}|\phi_{i}|^{2}+
\sum_{i,j}\chi^{+}_{i}\left[
{\hat\tau}_{0}\partial_{\tau}
+{\hat\tau}_{3}{i\over2}(\partial_{\tau}\theta_{j})
-{\hat\tau}_{3}\mu
-{\hat\tau}_{1}\Delta_{j}
\right]\delta_{i,j}\chi_{j}
\nonumber\\
&+&\sum_{i,j}\chi^{+}_{i}\left[
-t{\hat\tau}_{3}\sum_{\delta}\delta_{j,i+\delta}
e^{-i{\hat\tau}_{3}\theta_{i,j}/2}
\right]\chi_{j}\Bigg\}\;,
\eea
where $\theta_{i,j}=\theta_{i}-\theta_{j}$
is a phase difference between
sites $i$ and $j$, $\delta=\pm {\hat x}\;,
\pm{\hat y}$
for a square lattice.
Note that if  we equate $\phi_{j}=U C_{j\downarrow}C_{j\uparrow}$
and $\phi_{j}=\Delta_{j}e^{i\theta_{j}/2}$, then we know that
the phase part of $C_{j\sigma}$ is $\exp[i\theta_{j}/2]$. This means
that the gauge transformation\cite{gauge,kwon1} 
splits the charged fermion field $C_{j\sigma}$
into a charge bose field $\exp[i\theta_{j}/2]$ and a neutral fermion field
$\chi_{j}$, which obeys a Grassmann algebra.
 
After integrating
out fermion fields, we obtain $Z=\int\D\phi^{*}\D\phi \exp[-S_{eff}]$,
where the effective action is
\be
S_{eff}=\int d\tau\sum_{i}{1\over U}|\phi_{i}|^{2}-\mbox{Tr}\ln[G^{-1}]\;.
\ee
Here, $\mbox{Tr}$ means the trace over the functional space, depending on the 
representation, and on the spin space. We will use a symbol $\mbox{tr}$
for a trace over spin space only.
In the real (lattice) space, the Green function can be represented as
\be
G^{-1}(i,j)=G^{-1}_{0}(i,j)-\Sigma(i,j)\;,
\ee
where
\be
G^{-1}_{0}(i,j)=(-{\hat\tau}_{0}\partial_{\tau}+{\hat\tau}_{3}\mu
+{\hat\tau}_{2}\Delta)\delta_{i,j}+
{\hat\tau}_{3}
t\sum_{\delta}\delta_{j,i+{\delta}}
\ee
and the self energy has the form
\be
\Sigma(i,j)={i\over2}{\hat\tau}_{3}(\partial_{\tau}\theta_{j})\delta_{i,j}+
it{\hat\tau}_{0}
\sum_{\delta}\delta_{j,i+{\delta}}\sin\left({\theta_{i,j}\over2}\right)+
t{\hat\tau}_{3}\sum_{\delta}\delta_{j,i+{\delta}}
\left[1-\cos\left({\theta_{i,j}\over2}\right)\right].
\ee
\ewt
In the expression for $G^{-1}_{0}(i,j)$ the bulk value of $\Delta$ is used.
This means we consider only 
fluctuations of the phase and assume no fluctuation in the magnitude
of the gap.
Magnitude fluctuation will also change the self energy $\Sigma$. 
However, 
in this paper we concentrate on the effects of phase fluctuations.
For simplicity we also 
consider only the static case $(\partial_{\tau}\theta_{i}=0)$; therefore,
we are not concerned with the Landau damping effects which are related to
the scattering of the thermally excited quasiparticles with the excitations
of the phase field. However, if
one is interested in dynamics associated with the phase fluctuations,
then the Landau effects
have to be taken into account.\cite{aitchison}

Before going further we would like to
see what happens if the phase is treated in the 
continuum limit while the fermions are kept on a lattice.
Let us consider
$\sum_{i,j}\chi^{+}_{i}\Sigma(i,j)\chi_{j}$ in the static case.
Since we consider the continuum limit for the phase, we change the
phase difference to the phase derivative as follows:
\begin{equation}
\sum_{i,j}\chi^{+}_{i}{\hat\tau}_{0}it\sum_{\bf\delta}\delta_{j,i+{\bf\delta}}
\sin(\theta_{i,j}/2)\chi_{j}
\simeq\frac{1}{2}{\hat\tau}_{0}
it\sum_{i,\delta}\chi^{+}_{i}\theta_{i,i+\delta}\chi_{i+\delta}\;.
\end{equation}
Since $\theta_{i,i+\delta}=\theta_{i}-\theta_{i+\delta}\simeq
-\delta\cdot\nabla\theta-\frac{1}{2}(\delta\cdot\nabla)^{2}\theta$,
\bwt
\begin{equation}
\frac{1}{2}{\hat\tau}_{0}
it\sum_{i,\delta}\chi^{+}_{i}\theta_{i,i+\delta}\chi_{i+\delta}
\simeq
-\frac{1}{2}{\hat\tau}_{0}it\sum_{i,\delta}\chi^{+}_{i}
\left[\delta\cdot\nabla\theta+\frac{1}{2}(\delta\cdot\nabla)^{2}\theta\right]
\chi_{i+\delta}\;.
\end{equation}
The fermions are on the lattice so we need
to use Bloch state to expand
$\chi_{i}=\sum_{\bf k}e^{i{\bf k}\cdot{\bf r}_{i}}\chi_{\bf k}$.
Now we have for example
\be
\sum_{i,\delta}\chi^{+}_{i}(\delta\cdot\nabla\theta)\chi_{i+\delta}
=2i\sum_{{\bf k}}\chi^{+}_{\bf k}
\left[(\partial_{x}\theta)\sin(k_{x})+(\partial_{y}\theta)\sin(k_{y})\right]
\chi_{\bf k}\;.
\ee
Note that $\sum_{\delta}(\delta\cdot\nabla\theta)\cos({\bf k}\cdot\delta)=0$
because of symmetry.
We also have
\be
\sum_{i,\delta}\chi^{+}_{i}
\frac{1}{2}(\delta\cdot\nabla)^{2}\theta\chi_{i+\delta}
=\sum_{{\bf k}}\chi^{+}_{\bf k}
\left[(\partial^{2}_{x}\theta)\cos(k_{x})+(\partial^{2}_{y}\theta)\cos(k_{y})
\right]\chi_{\bf k}\;.
\ee
We therefore obtain
\begin{eqnarray}
&&\sum_{i,j}\chi^{+}_{i}{\hat\tau}_{0}it\sum_{\bf\delta}\delta_{j,i+{\bf\delta}}
\sin(\theta_{i,j}/2)\chi_{j}
\nonumber\\
\simeq&&
{\hat\tau}_{0}\sum_{{\bf k}}\chi^{+}_{\bf k}
\Bigg\{
-\frac{1}{2}it\left[(\partial^{2}_{x}\theta)\cos(k_{x})+(\partial^{2}_{y}\theta)
\cos(k_{y})
\right]
\nonumber\\
+&&t\left[(\partial_{x}\theta)\sin(k_{x})+
(\partial_{y}\theta)\sin(k_{y})\right]\Bigg\}\chi_{\bf k}\;.
\end{eqnarray}
Similarly, one can show that
\begin{eqnarray}
&&\sum_{i,j}\chi^{+}_{i}{\hat\tau}_{3}t
\sum_{\bf\delta}\delta_{j,i+{\bf\delta}}[1-\cos(\theta_{i,j}/2)]\chi_{j}
\nonumber\\
\simeq&&
\frac{1}{4}t{\hat\tau}_{3}\sum_{\bf k}\chi^{+}_{\bf k}
\left[(\partial_{x}\theta)^{2}\cos(k_{x})+(\partial_{y}\theta)^{2}\cos(k_{y})
\right]\chi_{\bf k}\;.
\end{eqnarray}
Consequently, when the phase is taken in the continuum limit while the fermions
are still on the lattice, the self energy in the momentum space
is
\begin{eqnarray}
\Sigma=&&
{\hat\tau}_{0}\left\{
-\frac{1}{2}it\left[(\partial^{2}_{x}\theta)\cos(k_{x})+(\partial^{2}_{y}\theta)
\cos(k_{y})
\right]+t\left[(\partial_{x}\theta)\sin(k_{x})+
(\partial_{y}\theta)\sin(k_{y})\right]\right\}
\nonumber\\
+&&\frac{1}{4}t
{\hat\tau}_{3}\left[(\partial_{x}\theta)^{2}\cos(k_{x})+(\partial_{y}\theta)^{2}
\cos(k_{y})
\right]\;,
\end{eqnarray}
which is the same as the self energy of Ref.\cite{sharapov}
\ewt

\section{effective action}

In this section we concentrate on the effective action for phase fluctuations.
By virtue of the gauge transformation used, the formalism associated with
the derivation of the desired action has become 
quite simple because the self-energy that we have obtained
in the previous section depends only on the phase of the order parameter. 

The effective action $S_{eff}$ can be separated into the mean-field
part $(S^{(0)})$ and the phase fluctuation part $(S_{\theta})$ as follows:
\bea
S_{eff}&=&\int d\tau\sum_{i}{1\over U}|\phi_{i}|^{2}-\mbox{Tr}\ln[G^{-1}]
\nonumber\\
&=&S^{(0)}-\mbox{Tr}\ln
\left[1-G_{0}\Sigma\right]
\nonumber\\
&=&S^{(0)}+S_{\theta}\;,
\eea
where
\be
S^{(0)}=\int d\tau\sum_{i}{1\over U}
|\Delta|^{2}-\mbox{Tr}\ln\left[G^{-1}_{0}\right]\;,
\ee
and
\be
S_{\theta}=\mbox{Tr}\sum_{n=1}^{\infty}
{1\over n}\left[G_{0}\Sigma\right]^{n}\;.
\ee
In the saddle-point approximation\cite{loktev}
$\partial S^{(0)}/\partial\Delta=0$ reduces to the BCS mean-field gap
equation:
\be
\Delta=\sum_{\bf k}{\Delta\over E_{\bf k}}
\tanh\left({E_{\bf k}\over2T}\right)
\label{gap}
\ee
and $-(T/V)(\partial S^{(0)}/\partial\mu)$,
where $V$ is a volume of the system, gives an equation for the filling
factor:
\be
n=1-\sum_{\bf k}{\xi_{\bf k}\over E_{\bf k}}
\label{filling}
\tanh\left({E_{\bf k}\over2T}\right)\;,
\ee 
where $E_{\bf k}=\sqrt{\xi^{2}_{\bf k}+\Delta^{2}}$ with
$\xi_{\bf k}=-2t\left[\cos(k_{x})+\cos(k_{y})\right]
-\mu$. 
In the calculation of 
the effective phase-only action $S_{\theta}$, 
we assumed that the phase fluctuations 
are small so that it is sufficient to consider
only the first and the second trace in the expansion for $S_{\theta}$; 
in other words,
$S_{\theta}\simeq S^{(1)}+S^{(2)}$, where
\be
S^{(1)}=\mbox{Tr}\left[G_{0}\Sigma\right]
\ee
and
\be
S^{(2)}={1\over2}\mbox{Tr}\left[G_{0}\Sigma G_{0}\Sigma\right]\;.
\ee

At finite temperature
we obtain an extended XY model Hamiltonian which includes not only
the nearest neighbor spin-spin interaction but also the 
next nearest neighbor and
third neighbor interaction. This constitutes an extension of the
usual XY Hamilton $(H_{XY})$, which contains
only nearest neighbor interactions.
Inclusion of the next nearest neighbor hopping 
in the electron energy dispersion makes the extended feature of the
effective Hamiltonian even more robust.
In this case the effective Hamiltonian manifests this extended property
even at zero temperature. This will be discussed in section V.

The easiest way to calculate $S_{\theta}$ is in
the momentum representation for $G_{0}$ and $\Sigma$. 
For simplicity, we will use a four vector notation: $K=({\bf k},\omega_{n})$,
$\sum_{K}=T\sum_{\omega_{n}}\sum_{\bf k}$, where $\omega_{n}$ is a 
fermionic Matsubara frequency, and $\int dX_{i}=\int d\tau\sum_{i}$
with the sum over the lattice sites.
It can be shown that
$\langle K|G|K'\rangle=\delta(K-K')G(K')$ and
$\langle K'|\Sigma|K\rangle={\hat\tau}_{0}{\tilde\Sigma}^{(0)}(K'-K,{\bf k})+
{\hat\tau}_{3}{\tilde\Sigma}^{(3)}(K'-K,{\bf k})$, where
\bwt
\be
{\tilde\Sigma}^{(0)}(K'-K,{\bf k})=
it\int dX_{i}e^{-i(K'-K)X_{i}}
\sum_{\bf \delta}e^{i{\bf k}\cdot{\bf \delta}}
\sin\left({\theta_{i,i+\delta}\over2}\right)\;,
\ee
and
\be
{\tilde\Sigma}^{(3)}(K'-K,{\bf k})=
t\int dX_{i}e^{-i(K'-K)X_{i}}
\sum_{\bf \delta}e^{i{\bf k}\cdot{\bf \delta}}
\left[1-\cos\left({\theta_{i,i+\delta}\over2}\right)\right]\;.
\ee
Since $S^{(1)}=\mbox{tr}\sum_{K,K'}\langle K|G_{0}|K'\rangle
\langle K'|\Sigma|K\rangle$, we obtain
\be
S^{(1)}=t\sum_{K}\mbox{tr}
\left[G(K){\hat\tau}_{3}e^{i\eta\omega_{n}{\hat\tau}_{3}}\right]
\cos(k_{x})\int d\tau\sum_{<ij>}\left[1-\cos\left({\theta_{i,j}\over2}
\right)\right],
\ee
\ewt
where $\eta\rightarrow0^{+}$ is a convergence factor,\cite{abrikosov}
which originates
from $G(0,\tau-\tau^{+})$. Note that $\Sigma^{(0)}$ makes no contribution
to $S^{(1)}$. Since $S^{(1)}$ include the phase difference between
the nearest neighbor sites, 
the first non-trivial term is $\theta^{2}_{i,j}$. We
obtain immediately the action of the usual XY model; 
however, a $\theta^{2}_{i,j}$ term
will also appear in $S^{(2)}$. Moreover, the phase difference between
the next nearest neighbor sites is also 
obtained from the second trace. As we will see later Section V,
if we include the next nearest neighbor hopping, then even $S^{(1)}$ contains 
the 
phase difference between next nearest neighbor sites.

The second trace can also be calculated in a similar manner even though
it is more complicated:
\bwt
\be
S^{(2)}={1\over2}\mbox{tr}\sum_{K,K'}G(K){\tilde\Sigma}(K-K',{\bf k}')
G(K'){\tilde\Sigma}(K'-K,{\bf k})\;, 
\ee
where
${\tilde\Sigma}={\hat\tau}_{0}{\tilde\Sigma}^{(0)}
+{\hat\tau}_{3}{\tilde\Sigma}^{(3)}$.
After some manipulation, one can arrive at
\bea
S^{(2)}=&&{1\over2}\sum_{K,K'}\mbox{tr}\left[
G(K)G(K')\right]\tilde{\Sigma}^{(0)}(K-K',{\bf k}')
\tilde{\Sigma}^{(0)}(K'-K,{\bf k})
\nonumber\\
+&&{1\over2}\sum_{K,K'}\mbox{tr}\left[
G(K){\hat\tau}_{3}G(K'){\hat\tau}_{3}\right]\tilde{\Sigma}^{(3)}(K-K',{\bf k}')
\tilde{\Sigma}^{(3)}(K'-K,{\bf k}).
\eea
It can be shown that the mixed terms do not contribute to $S^{(2)}$
in our consideration which deals only with a local effective action.
In order to calculate $S^{(2)}$ we need further manipulations. Define
$A_{\delta}(X_{i})=\sin(\theta_{i,i+\delta}/2)$ and express it in terms of
its momentum counterpart as follows: 
$A_{\delta}(X_{i})=\sum_{Q}{\tilde A}_{\delta}(Q)\exp[iQX_{i}]$, where
$Q$ is a four vector $({\bf q}, \Omega_{m})$ with a bosonic Matsubara
frequency $\Omega_{m}$. 
Then, for example, one can show for the first term of $S^{(2)}$ that
\bea
&&\sum_{K,K'}\mbox{tr}\left[
G(K)G(K')\right]\tilde{\Sigma}^{(0)}(K-K',{\bf k}')
\tilde{\Sigma}^{(0)}(K'-K,{\bf k})
\nonumber\\
=&&(it)^{2}\sum_{\delta,\delta'}\sum_{Q}{\tilde A}_{\delta}(Q)
{\tilde A}_{\delta'}(-Q)e^{-i{\bf q}\cdot\delta}\sum_{K}
\mbox{tr}\left[G(K)G(K-Q)\right]e^{i{\bf k}\cdot(\delta+\delta')}\;.
\eea
\ewt
See Appendix A for a detailed derivation.
A similar expression can be obtained for the second term.
Since, as we mentioned earlier, we are concerned only with the local effective
action for the phase only and since we wish to 
see if the equivalence between
the phase-only Hamiltonian and the usual
$H_{XY}$, which is found in the continuum limit, 
still holds on a lattice, we expand $S^{(2)}$ about $Q=0$ and keep the
leading order. 
Here we would like to explain that in contrast
with the continuum limit this procedure requires a periodic lattice.
In the continuum limit ${\tilde A}_{\delta}(Q)
{\tilde A}_{\delta'}(-Q)\sim q^{2}\theta_{q}\theta_{-q}$; therefore,
we can put $Q=0$ for $\mbox{tr}\left[G(K)G(K-Q)\right]$.
However, this is not the case on the lattice
because $A_{\delta}(X_{i})$ does not deal with the phase derivative but
rather it deals with the
phase difference. One might expand ${\tilde A}_{\delta}(Q)$
and $\mbox{tr}\left[G(K)G(K-Q)\right]$
in terms of $Q$ and, then, regroup terms of the same order of $Q$.
If we do so 
however, we cannot transform back to the lattice space with
a function of the phase difference.\cite{bc} 
Instead of doing so, we see under what condition we may simply put
$Q=0$ for $\mbox{tr}\left[G(K)G(K-Q)\right]$  and neglect corrections. 
What we
found is a periodic boundary condition for the phase $\theta_{i}$.
If the size of the 2D lattice is $L\times L$, then
$\theta({\bf r}_{i}+L)=\theta({\bf r}_{i})$. 
See Appendix B for details.

Now, it can be shown that
\bwt
\bea
S^{(2)}&=&
{1\over2}t^{2}\sum_{\delta,\delta'}\sum_{K}\mbox{tr}\left[G(K)G(K)\right]
e^{i{\bf k}\cdot(\delta-\delta')}\int dX_{i}
\sin\left({\theta_{i,i+\delta}\over2}\right)
\sin\left({\theta_{i,i+\delta'}\over2}\right)
\nonumber\\
&+&
{1\over2}t^{2}\sum_{\delta,\delta'}\sum_{K}\mbox{tr}
\left[G(K){\hat\tau}_{3}G(K){\hat\tau}_{3}\right]
e^{i{\bf k}\cdot(\delta-\delta')}
\nonumber\\
&\times&
\int dX_{i}
\left[1-\cos\left({\theta_{i,i+\delta}\over2}\right)\right]
\left[1-\cos\left({\theta_{i,i+\delta'}\over2}\right)\right]\;.
\eea
\ewt
In order to
compare the effective Hamiltonian with the $H_{XY}$, we need to expand
$S^{(2)}$ in terms of the phase differences while keeping terms up to
$\theta^{2}_{i,j}$. Since the first non-trivial term for the second 
term is $\theta^{4}_{i,j}$,
we will ignore it in the effective Hamiltonian.
Using symmetry properties for $G(K)$ such as remaining the same with respect to
the exchange of $k_{x}$ and $k_{y}$, we obtain the effective 
local Hamiltonian: $\H_{\theta}=\langle\theta|{\hat M}|\theta\rangle$,
where $\langle\theta|=(\theta_{i,i+{\hat x}},\theta_{i,i-{\hat x}},
\theta_{i,i+{\hat y}},\theta_{i,i-{\hat y}})$ and 
\begin{center}
${\hat M}=\left(\begin{array}{cccc}
         \alpha&\beta&\gamma&\gamma\\
         \beta&\alpha&\gamma&\gamma\\
         \gamma&\gamma&\alpha&\beta\\
         \gamma&\gamma&\beta&\alpha
\end{array}\right)$
\end{center}
with components as follow:
\bwt
\bea
\alpha&=&{1\over8}t\sum_{K}\mbox{tr}
\left[G(K){\hat\tau}_{3}e^{i\eta\omega_{n}{\hat\tau}_{3}}\right]
\cos(k_{x})+{1\over8}t^{2}\sum_{K}\mbox{tr}\left[G(K)G(K)\right]
\\
\beta&=&{1\over8}t^{2}\sum_{K}\mbox{tr}\left[G(K)G(K)\right]\cos(2k_{x})\;,
\eea
\ewt
and
\be
\gamma={1\over8}t^{2}\sum_{K}
\mbox{tr}\left[G(K)G(K)\right]\cos(k_{x})\cos(k_{y})\;.
\ee
As one can see, the $4\times4$ matrix ${\hat M}$ is not diagonal. 
This means
that the effective Hamiltonian $\H_{\theta}$ is not equivalent to
the usual $H_{XY}$, which would be pure-diagonal.
Not only do we have terms like $\theta^{2}_{i,i+{\hat x}}$ and
$\theta^{2}_{i,i-{\hat x}}$ but we also have terms of the form
$\theta_{i,i+{\hat x}}\theta_{i,i-{\hat x}}$.
See Appendix A.
We will investigate effects of these off-diagonal
terms in the next section.

It is worthwhile making sure our new result for $\H_{\theta}$ reduces
to the well known XY-type Hamiltonian\cite{loktev} in the continuum limit.
For this purpose, we need to recover the explicit dependence on the
lattice constant $a$ 
in the expression
for $\H_{\theta}$
which we have set equal to $1$. This will allow us
to track orders of $a$ and take the limit $a\rightarrow0$.
In this case
the phase difference becomes a derivative of the phase $(\nabla\theta)$
while $ta^{2}\rightarrow{1\over 2m}$, where
$m$ is an effective mass of the electron. It is straightforward to show
that, in the continuum limit with $\xi_{\bf k}={k^{2}\over 2m}-\mu$,
$\H_{\theta}\rightarrow H^{(1)}_{\theta}+H^{(2)}_{\theta}$, where
\be
H^{(1)}_{\theta}={1\over 8m}\sum_{K}\mbox{tr}
\left[G(K){\hat\tau}_{3}e^{i\eta\omega_{n}{\hat\tau}_{3}}\right]
\int d{\bf r}\left(\nabla\theta\right)^{2}\;,
\ee
and
\be
H^{(2)}_{\theta}={1\over 16m^{2}}\sum_{K}\mbox{tr}\left[G(K)G(K)\right]k^{2}
\int d{\bf r}\left(\nabla\theta\right)^{2}\;.
\ee
See Appendix C for details.
Consequently, the effective Hamiltonian $\H_{\theta}$ we have derived
does reduce to
the well-known XY-type expression,
$H_{XY}=J_{XY}\int d{\bf r}\left(\nabla\theta\right)^{2}$. 

\begin{figure}[tp]
\begin{center}
\includegraphics[height=2.6in,width=3in]{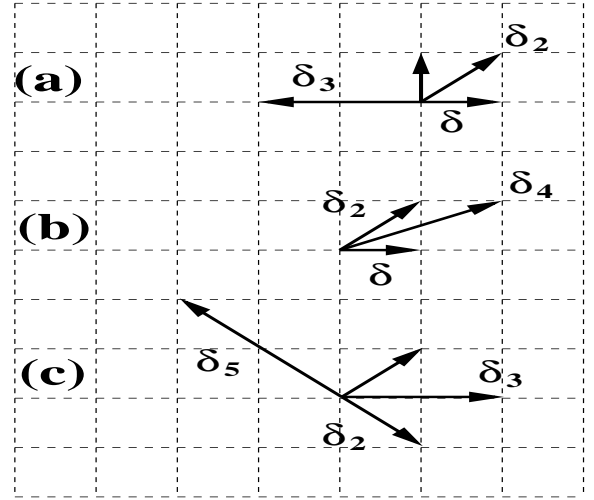}
\caption{Geometrical illustrations of induced interactions
for spin pairs. $\delta$ is a vector to the nearest neighbor site.
$\delta_{2}$ is to the next nearest neighbor site and so forth.
}
\end{center}
\end{figure}

\section{off-diagonal terms}

Let us now consider effects of the off-diagonal terms in $\H_{\theta}$.
At $T=0$, $\H_{\theta}$ again becomes equivalent to the XY-type Hamiltonian
because the off-diagonal terms of ${\hat M}$, namely 
$\beta$ and $\gamma$ vanish.
However, at finite temperature they are finite so that 
$\H_{\theta}$ is no longer of the usual XY type; that is
\bea
\H_{\theta}=&&\alpha\sum_{<ij>}\theta^{2}_{i,j}
+2\beta\sum_{i}\left(
\theta_{i,i+{\hat x}}\theta_{i,i-{\hat x}}+
\theta_{i,i+{\hat y}}\theta_{i,i-{\hat y}}\right)
\nonumber\\
+&&2\gamma\sum_{i}
\left(\theta_{i,i+{\hat x}}+\theta_{i,i-{\hat x}}\right)
\left(\theta_{i,i+{\hat y}}+\theta_{i,i-{\hat y}}\right)\;.
\eea
Since the phase fluctuation between two sites is small by
assumption,
one can make the approximation that $\theta_{i,i+{\hat x}}\simeq
\sin(\theta_{i,i+{\hat x}})$. Introducing a 2D classical spin 
${\bf S}_{i}=(\cos(\theta_{i}),\sin(\theta_{i}))$ 
at a site $i$, it can be shown, within the approximation
we have made, that,
for example,
$\theta_{i,i+{\hat x}}\theta_{i,i-{\hat x}}\simeq
({\bf S}_{i}\times{\bf S}_{i+{\hat x}})\cdot
({\bf S}_{i}\times{\bf S}_{i-{\hat x}})$.
It can be seen that a spin
at the site $i+{\hat x}$ couples to a spin at $i-{\hat x}$, which
is the next nearest neighbor spin-spin interaction. The only assumption
we make is that terms higher than $\theta^{2}_{i,j}$ for
a given $(i,j)$ are negligible.
This does not mean, however, that only the nearest neighbor phase 
differences are important. 
Following the procedure we described in Appendix D, one can show that
the second term of $\H_{\theta}$, which is proportional to $\beta$, becomes
$\beta\sum_{<ij>}\theta^{2}_{i,j}+\beta\sum_{<ij>_{3}}{\bf S}_{i}\cdot
{\bf S}_{j}$ up to a constant, 
where the symbol $<ij>_{3}$ indicates a sum over the next-next
nearest pairs.
Similarly, the third term of $\H_{\theta}$ proportional to $\gamma$ turns
out to be
$2\gamma\sum_{<ij>}\theta^{2}_{i,j}
+2\gamma\sum_{\ll ij\gg}{\bf S}_{i}\cdot{\bf S}_{j}$,
where $\ll ij\gg$ means a sum over the next nearest pairs. 
Since
$\theta^{2}_{i,j}$ can also be represented in term of
${\bf S}_{i}\cdot{\bf S}_{j}$, the effective
Hamiltonian $\H_{\theta}$ can be written as
\be
\H_{\theta}=-J_{1}\sum_{<ij>}{\bf S}_{i}\cdot{\bf S}_{j}
+J_{2}\sum_{\ll ij\gg}{\bf S}_{i}\cdot{\bf S}_{j}
+J_{3}\sum_{<ij>_{3}}{\bf S}_{i}
\cdot{\bf S}_{j}\;,
\label{htheta}
\ee
where $J_{1}=2(\alpha+\beta+2\gamma)$, $J_{2}=2\gamma$, and
$J_{3}=\beta$.
It is clear that this Hamiltonian is not of the usual XY type
but instead of an extended XY type.  
A geometrical explanation for the appearance of the next nearest neighbor
and the next-next nearest neighbor term in Eq.~(\ref{htheta})
is illustrated
in Fig.~1(a). Indeed, these terms come from the second trace
proportional to $t^{2}$, which has a factor
${\tilde\Sigma}^{(0)}{\tilde\Sigma}^{(0)}$. Each self-energy 
${\tilde\Sigma}^{(0)}$ picks up ${\bf \delta}=\pm{\hat x},\;\pm{\hat y}$,
and the second trace gives terms with resulting vectors
$\delta+\delta'=\delta_{2}$ or $\delta_{3}$,
where $\delta_{2}=\pm{\hat x}\pm{\hat y}$ and
$\delta_{3}\pm2{\hat x},\;\pm2{\hat y}$. 
This geometrical picture
also works when we include the next nearest neighbor hopping 
$(t')$ in the electron
dispersion curves.

The physics of $\H_{\theta}$ depends on the relative magnitudes of
coefficients $J_{1}$, $J_{2}$, and $J_{3}$ in Eq.~(\ref{htheta})
as well as their relative signs.
For example, if $\beta$ is negligible, and $\alpha$ and $\gamma$ are both
positive so that $J_{1}>2J_{2}$,
$\H_{\theta}$ describe a non-frustrated XY model, and its
critical behavior can be understood in terms of the usual XY Hamiltonian
with an effective coupling constant $J_{eff}=(J_{1}-2J_{2})$.\cite{simon}
However, this does not mean that in this case
$\H_{\theta}$ is equivalent to
$-J_{eff}\sum_{<ij>}{\bf S}_{i}\cdot{\bf S}_{j}$
because
the local behavior of these two Hamiltonians is different.
In general, however, as long as $J_{1}$ is dominant, the large length scale
behavior is of the usual XY type.
To calculate $J_{1}$, $J_{2}$, and $J_{3}$,
we need to know how $\Delta$ and $\mu$ change with increasing
temperature.
We choose the pairing interaction $U=1.4t$ and the filling factor
$n=0.9$ and self-consistently solve Eqs. (\ref{gap}) and (\ref{filling})
to determine $\Delta(T)$ and $\mu(T)$ as functions of $T$.
In Fig.~2 we plot $J_{eff}$ vs $T$ scaled by $T_{MF}$
for $t'=0$.
The value of the BKT transition temperature $(T_{BKT})$
is indicated by an arrow. As one can see, in this case
$T_{BKT}/T_{MF}\simeq0.73$. If $n$ gets closer to $1$ as well as
$U$ becomes smaller; namely, $n\approx 1$ and $U < t$,
$T_{BKT}/T_{MF}\approx 1$ as we verified numerically.

\vskip 1.5cm
\begin{figure}[tp]
\begin{center}
\includegraphics[height=2.6in,width=3in]{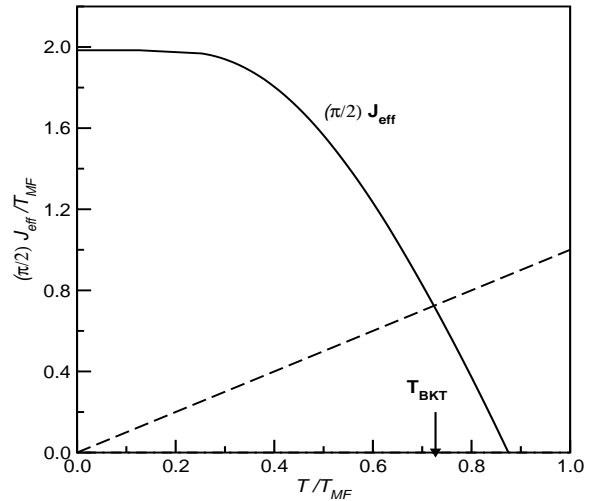}
\caption{
$\frac{\pi}{2}{\cal J}_{eff}$ (solid curve) as a function of temperature $(T)$
scaled by $T_{MF}$ for the pairing interaction $U=1.4t$ and
the filling factor $n=0.9$ with
next nearest neighbor hopping $t'=0$.
The solution of $T=\frac{\pi}{2}{\cal J}_{eff}(T)$ gives
the BKT temperature $T_{BKT}$ which is indicated as an arrow.
}
\end{center}
\end{figure}

\section{inclusion of the next-nearest hopping}

When we include the next nearest neighbor hopping $(t')$, the Hamiltonian
becomes
\bwt
\be
\H=-t\sum_{<ij>\sigma}C^{+}_{i\sigma}C_{j\sigma}
-t'\sum_{\ll ij\gg\sigma}C^{+}_{i\sigma}C_{j\sigma}
-\mu\sum_{i\sigma}n_{i\sigma}
-U\sum_{i}n_{i\uparrow}n_{i\downarrow}\;,
\ee
where $\ll ij\gg$ denotes a summation over the next nearest neighbor sites, and
now $\xi_{\bf k}=-2t\left[\cos(k_{x})+\cos(k_{y})\right]
-4t'\cos(k_{x})\cos(k_{y})
-\mu$.
The same procedure as before yields the actions
\bea
S&=&\int d\tau\Bigg\{
{1\over U}\sum_{i}|\phi_{i}|^{2}+
\sum_{i,j}\chi^{+}_{i}\left[
{\hat\tau}_{0}\partial_{\tau}
+{\hat\tau}_{3}{i\over2}(\partial_{\tau}\theta_{j})
-{\hat\tau}_{3}\mu
-{\hat\tau}_{1}\Delta_{j}
\right]\delta_{i,j}\chi_{j}
\nonumber\\
&+&\sum_{i,j}\chi^{+}_{i}\left[
-t{\hat\tau}_{3}\sum_{\delta}\delta_{j,i+\delta}
e^{-i{\hat\tau}_{3}\theta_{i,j}/2}
-t'{\hat\tau}_{3}\sum_{\delta}\delta_{j,i+\delta_2}
e^{-i{\hat\tau}_{3}\theta_{i,j}/2}
\right]\chi_{j}\Bigg\}\;,
\eea
and the Green function
\be
G^{-1}_{0}(i,j)=(-{\hat\tau}_{0}\partial_{\tau}+{\hat\tau}_{3}\mu
+{\hat\tau}_{2}\Delta)\delta_{i,j}+
{\hat\tau}_{3}\left[t\sum_{\bf\delta}\delta_{j,i+{\bf\delta}}+
t'\sum_{\delta_{2}}\delta_{j,i+{\bf\delta}_{2}}\right]\,
\ee
with the self energies
\be
\Sigma^{(0)}(i,j)=
it\sum_{\bf\delta}\delta_{j,i+{\bf\delta}}
\sin\left({\theta_{i,j}\over2}\right)+
it'\sum_{\delta_{2}}\delta_{j,i+{\bf\delta}_{2}}
\sin\left({\theta_{i,j}\over2}\right)\;,
\ee
and
\be
\Sigma^{(3)}(i,j)=t\sum_{\bf\delta}\delta_{j,i+{\bf\delta}}
\left[1-\cos\left({\theta_{i,j}\over2}\right)\right]+
it'\sum_{\delta_{2}}\delta_{j,i+{\bf\delta}_{2}}
\left[1-\cos\left({\theta_{i,j}\over2}\right)\right]\;.
\ee
We obtain the effective Hamiltonian including
the next nearest neighbor hopping contribution:
\be
\H_{\theta}=-{\cal J}_{1}\sum_{<ij>}{\bf S}_{i}\cdot{\bf S}_{j}
+{\cal J}_{2}\sum_{\ll ij\gg}{\bf S}_{i}\cdot{\bf S}_{j}
+{\cal J}_{3}\sum_{<ij>_{3}}{\bf S}_{i}\cdot{\bf S}_{j}
+{\cal J}_{4}\sum_{<ij>_{4}}{\bf S}_{i}\cdot{\bf S}_{j}
+{\cal J}_{5}\sum_{<ij>_{5}}{\bf S}_{i}\cdot{\bf S}_{j}\;,
\ee
\ewt
where $\sum_{<ij>_{4(5)}}$ means $\sum_{i,\delta_{4(5)}}$.
Figs.~1(b) and (c) show
geometrical descriptions for theses terms.
In the same way that the $t^{2}$ term gives interactions of the form
${\bf S}_{i}\cdot{\bf S}_{i+\delta_{2}}$ and
${\bf S}_{i}\cdot{\bf S}_{i+\delta_{3}}$, the $tt'$ term induces the form
${\bf S}_{i}\cdot{\bf S}_{i+\delta_{4}}$ because
$\delta+\delta_{2}=\delta_{4}$, and the $t'^{2}$ term gives rise to
${\bf S}_{i}\cdot{\bf S}_{i+\delta_{3}}$ and
${\bf S}_{i}\cdot{\bf S}_{i+\delta_{5}}$
because $\delta_{2}+\delta'_{2}=\delta_{3}$ or $\delta_{5}$, where
$\delta_{4}=\pm{\hat x}\pm2{\hat y},\;\pm2{\hat x}\pm{\hat y}$, and
$\delta_{5}=\pm2{\hat x}\pm2{\hat y}$. 

It is obvious that the extended features of $\H_{\theta}$ are
robust in band structure parameters. 
Moreover, contrary to what we found before, now 
${\cal J}_{2}$ is finite even at $T=0$ because
the $t'$-term contributes to the first trace as well as
to the second trace as shown in Appendix E, where detailed expressions
for ${\cal J}_{i}\; (i=1,2,3,4,5)$ are presented. In order to calculate
the ${\cal J}_{i}$'s we choose $t'=-0.2t$ for several values 
of parameters $(U,n)$. In Fig.~3, from the top to bottom
panel $(U,n)$ are $(1.4,0.9)$,
$(2.0,0.9)$ and $2.0,0.4)$, respectively.
Since our numerical calculation indicates that
${\cal J}_{1}$ and ${\cal J}_{2}$ are dominant and 
${\cal J}_{1}>2{\cal J}_{2}$ in most of the temperature range, we neglect 
${\cal J}_{3}$, ${\cal J}_{4}$, and ${\cal J}_{5}$. 
In this case the critical behavior
is still of the usual XY type so that we introduce an effective coupling
${\cal J}_{eff}={\cal J}_{1}-2{\cal J}_{2}$. 
As also shown in Fig.~3, $(\pi/2){\cal J}_{eff}/T_{MF}$
becomes smaller as $U$ is increased or $n$ is decreased; therefore,
the crossing point between $(\pi/2){\cal J}_{eff}/T_{MF}$ and
$T/T_{MF}$ changes correspondingly.
For the attractive Hubbard model, the $(U,n)$ phase diagram
shows BCS, cross-over, and BE regimes; for example,
a large $n(\approx1)$ and a small $U(\ll t)$ corresponds to the BCS regime
while a small $n$ and a large $U$ corresponds to the BE regime.\cite{andrenacci}
Consequently, in a qualitative sense $T_{BKT}/T_{MF}$ becomes smaller
as the parameter moves from the BCS to the BE regime. 

\begin{figure}[tp]
\begin{center}
\includegraphics[height=2.6in,width=3in]{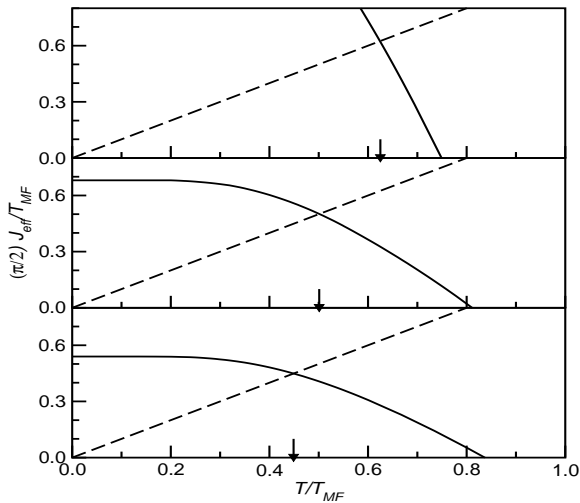}
\caption{
$\frac{\pi}{2}{\cal J}_{eff}$ (solid curves) as functions of temperature
for various values of
the pairing interaction $U$ and the filling factor $n$.
From the top to bottom $(U,n)$ are
$(1.4,0.9)$,
$(2.0,0.9)$ and $2.0,0.4)$, respectively, with $t'=0.2t$.
The BKT temperature $T_{BKT}$ is indicated as an arrow in each panel
It is given by the intersection of the solid curve for
$\frac{\pi}{2}{\cal J}_{eff}$ and the dashed line.
$(\pi/2){\cal J}_{eff}/T_{MF}$
becomes smaller as $U$ is increased or $n$ is decreased; therefore,
the crossing point between $(\pi/2){\cal J}_{eff}/T_{MF}$ and
$T/T_{MF}$ changes correspondingly.
As we move away from weak BCS regimes toward strong coupling
the size of the phase fluctuation regime between the mean field temperature
$(T^{*})$ and the BKT temperature $(T_{c})$ increases.
}
\end{center}
\end{figure}

\section{discussions and conclusions}

An effective Hamiltonian for phase fluctuations has been derived
on a square lattice using the attractive Hubbard model to describe
electron dynamics.
We find that the effective Hamiltonian is not
of the usual XY type, which maps onto a spin Hamiltonian with nearest
neighbors only,
but is extended XY containing spin interaction
between second up to fifth neighbors. 
This is in contrast to the
common assumption that discretization of the continuum limit Hamiltonian
leads to its lattice version.
In the continuum limit the BCS phase stiffness and hence the effective
Hamiltonian
vanish at $T_{MF}$ because they are proportional to the number of
Cooper pairs. In other words $T_{MF}$ can be determined not only
from the mean field equation for the gap but also from the phase stiffness
in the continuum limit. However, this is not the case on the lattice.
Since we showed that the effective Hamiltonian we obtained 
reduces properly to the well-known continuum limit result,
this is clearly a special property of the
continuum limit and does not hold on a lattice. 
It is interesting to note that the periodic boundary conditions in
our formalism are essential to get simple results and they are
those that are usually assumed for summation over the \bz.
Without them the effective Hamiltonian would not necessarily
end at a finite number of neighbors.

The extended feature of the effective phase Hamiltonian is reinforced
when the next nearest neighbor hopping is considered in the electronic
dispersion.
The critical behavior is however still BKT-like but with an
effective coupling which depends significantly
on temperature. 
This arises because near the BKT transition the extended Hamiltonian reduces
to the XY case but this simplification only holds in this restricted region of
temperature. In particular the generalized phase Hamiltonian does not vanish
at the mean field temperature $T_{MF}$ which is identified with the \pg
temperature $T^{*}$.
The crossover from BCS to Bose-Einstein
corresponding to increasing $U$ and/or decreasing $n$ reduces
the ratio of the BKT transition temperature to the mean
field pseudogap temperature.

In the continuum limit, $1-T_{BKT}/T_{MF}=4T_{MF}/E_{F}$, where $E_{F}$
is the Fermi energy, for the BCS limit (large $n$ and small $U$).\cite{loktev}
Since $T_{MF}/E_{F}$ is negligibly small for typical
values of $T_{MF}\approx 20-30$ meV and $E_{F}\approx 800-1000$ meV, 
the temperature region between $T_{BKT}(\equiv T_{c})$ and 
$T_{MF}(\equiv T^{*})$ is negligibly small as well. 
However, on the lattice we find instead $1-T_{BKT}/T_{MF}=0.375$
for $n=0.9$ and $U=1.4$, which are typical values of these 
parameters in the BCS
regime. Beyond the BCS regime (small $n$ and large $U$),
$T_{BKT}/T_{MF}$ is mostly
$0.5$ in the continuum limit while the ratio
reduces to $0.22$ for $n=0.4$ and $U=2.8$ on the lattice. Therefore,
the range for phase fluctuations between $T_{c}$ and $T^{*}$
can be much larger on the
lattice as compared with the continuum limit.

In our consideration $J_{1}$ (or ${\cal J}_{1}$ for $t'\ne0$) 
is not the BCS phase
stiffness. Thus it does not have to vanish at $T_{MF}$.
However, $J_{eff}$ (or ${\cal J}_{eff}$) is the phase stiffness near the
critical region and it determines $T_{BKT}$. Nevertheless
we want to mention 
that we cannot interpret $J_{eff}$ as the BCS stiffness
in the whole temperature range. It is only near $T_{BKT}$
that $-J_{eff}\sum_{<i,j>}{\bf S}_{i}\cdot{\bf S}_{j}$ can represent
the full Hamiltonian ${\cal H}_{\theta}$ that we obtained on the lattice.

We have considered only $s$-wave pairing in this paper. 
For cuprates
$d$-wave pairing is appropriate and this case
received considerable attention in the literature but only
in the continuum limit.\cite{kwon1,sharapov,paramekan,kwon2,benfatto,mitrovic} 
In the continuum limit the order parameter $\phi({\bf R},{\bf r})$,
where ${\bf R}({\bf r})$ is the center of mass (relative) coordinate,
is represented by $\phi({\bf R},{\bf r})=
\Delta({\bf R},{\bf r})e^{i\theta({\bf R})}$ because the bond phase is
usually replaced by the site phase. However, on the lattice 
it is the bonding order between the nearest sites that we must use 
as the building block to construct the $d$-wave order parameter.
This means that we can consider phase fluctuations between the
nearest neighbor building blocks
in addition to the phase difference by $\pi$ between the two;
therefore, we may expect that a subdominant component to the order
parameter may be induced by phase fluctuations.

We neglected the so-called Landau terms
associated with damping effects in this work. These terms do not enter
in the static case which we have considered here.
For dynamics, the Landau damping effects may play a role. Nevertheless
these effects have been shown\cite{aitchison} 
not to be important for $T\lesssim 0.6T_{MF}$
in the case of
$s$-wave superconductor in the continuum limit. 
However, it has also been pointed out that
because of the nodal structure of the order parameter of 
a $d$-wave superconductor
the Landau terms 
can have important effects even at low temperature.\cite{sharapov}
Therefore, to describe the dynamics of the phase fluctuations in a $d$-wave 
superconductor one has to consider Landau damping effects. 
Such effects will also be important on a lattice.
An extension of the present work to include Landau
terms is necessary if dynamics is considered.
Further complications associated with the existence of nodal quasiparticles
have been noted in Ref.\cite{mitrovic}

\begin{acknowledgments}
We acknowledge S. Sharapov for stimulating and elucidating discussions.
This work was supported in part by the Natural Sciences and Engineering
Research Council of Canada (NSERC), by ICORE (Alberta), 
and by the Canadian Institute for Advanced
Research (CIAR).
\end{acknowledgments}

\appendix
\section{manipulations for the second trace}

In this Appendix we manipulate the second trace. Let us consider
\bwt
\be
\sum_{K,K'}\mbox{tr}\left[
G(K)G(K')\right]\tilde{\Sigma}^{(0)}(K-K',{\bf k}')
\tilde{\Sigma}^{(0)}(K'-K,{\bf k})\;,
\label{a1}
\ee
where
\be
\tilde{\Sigma}^{(0)}(K-K',{\bf k}')=
it\int dX_{i}e^{-i(K-K')X_{i}}\sum_{\delta}e^{i{\bf k}'\cdot\delta}
A_{\delta}(X_{i})
\ee
with
\be
A_{\delta}(X_{i})=\sin\left({\theta_{i,i+\delta}\over2}\right)
=\sum_{Q}{\tilde A}_{\delta}(Q)e^{iQX_{i}}\;.
\ee
Now Eq.~(\ref{a1}) become
\bea
&&\sum_{K,K'}\mbox{tr}\left[G(K)G(K')\right]
\int dX_{i}dX_{j}e^{-i(K-K')(X_{i}-X_{j})}
\nonumber\\
&&\times (it)^{2}\sum_{\delta,\delta'}
e^{i{\bf k}'\cdot\delta+i{\bf k}\cdot\delta'}
\sum_{Q,Q'}{\tilde A}_{\delta}(Q)e^{iQX_{i}}
{\tilde A}_{\delta'}(Q')e^{iQ'X_{j}}
\nonumber\\
=&&(it)^{2}\sum_{\delta,\delta'}\sum_{Q}{\tilde A}_{\delta}(Q)
{\tilde A}_{\delta'}(-Q)e^{-i{\bf q}\cdot\delta}
\sum_{K}\mbox{tr}\left[G(K)G(K-Q)\right]
e^{i{\bf k}\cdot(\delta+\delta')}
\eea

For the local effective action, we put $Q=0$ in
$\mbox{tr}\left[G(K)G(K-Q)\right]$, which means that
we keep only the leading term in the expansion of Eq.~(\ref{a1}). 
\cite{loktev,aitchison,sharapov}
Note that $\sin(\theta_{i,i+\delta})\simeq \theta_{i,i+\delta}
\sim\nabla\theta\sim q{\tilde \theta}(q)$ so that
$\sin(\theta_{i,i+\delta})\sin(\theta_{i,i+\delta'})\sim
q^{2}{\tilde \theta}(q){\tilde \theta}(-q)$. Consequently,
$\mbox{tr}\left[G(K)G(K-Q)\right]$ with a finite $Q$
gives higher order terms in the expansion.
Since 
\be
\sum_{Q}{\tilde A}_{\delta}(Q)
{\tilde A}_{\delta'}(-Q)e^{-i{\bf q}\cdot\delta}
=\int dX_{i}{\tilde A}_{\delta}(i-\delta,\tau)
{\tilde A}_{\delta'}(i,\tau)\;,
\ee
we have for Eq.~(\ref{a1})
\be
t^{2}\sum_{\delta,\delta'}
\sum_{K}\mbox{tr}\left[G(K)G(K)\right]
e^{i{\bf k}\cdot(\delta-\delta')}
\int dX_{i}\sin\left({\theta_{i,i+\delta}\over2}\right)
\sin\left({\theta_{i,i+\delta'}\over2}\right)\;.
\ee
The manipulation for the second term exactly parallels the above
derivation. However, as we can see, the second term is negligible
because of the assumption of slow variation of the phase; 
therefore, we neglect it.

Now we have an effective Hamiltonian:
\bea
\H_{eff}&=&
t\sum_{K}\mbox{tr}
\left[G(K){\hat\tau}_{3}e^{i\eta\omega_{n}{\hat\tau}_{3}}\right]
\cos(k_{x})\tau\sum_{<ij>}\left[1-\cos\left({\theta_{i,j}\over2}
\right)\right]
\nonumber\\
&+&
{1\over2}t^{2}\sum_{\delta,\delta'}\sum_{K}\mbox{tr}\left[G(K)G(K)\right]
e^{i{\bf k}\cdot(\delta-\delta')}\sum_{i}
\sin\left({\theta_{i,i+\delta}\over2}\right)
\sin\left({\theta_{i,i+\delta'}\over2}\right)
\eea
For a small variation of $\theta_{i,j}$, one can derive
an effective local Hamiltonian $\H_{\theta}$.
Its derivation is straightforward so that we only briefly mention
some details about it. 
For the second term of $\H_{eff}$, if $\delta=\delta'$, then
one has ${1\over4}\mbox{tr}\left[G(K)^{2}\right]\sum_{i,\delta}
\left(\theta_{i}-\theta_{i+\delta}\right)^{2}$.
When $\delta\ne\delta'$, for example $\delta={\hat x}$ and
$\delta'=-{\hat x}$, one has
${1\over4}\mbox{tr}\left[G(K)^{2}\right]\cos(2k_{x})
\sum_{i}\left(\theta_{i}-\theta_{i+{\hat x}}\right)
\left(\theta_{i}-\theta_{i-{\hat x}}\right)$.
For other cases, one can apply similar manipulation. Note that
$\delta$ and $\delta'$ are exchangeable.

\section{boundary condition}

Let us explain how we end up with the boundary condition.
Consider
\begin{eqnarray}
&&\sum_{\delta,\delta'}\sum_{Q}{\tilde A}_{\delta}(Q)
{\tilde A}_{\delta'}(-Q)e^{-i{\bf q}\cdot\delta}\sum_{K}
e^{i{\bf k}\cdot(\delta+\delta')}
\Lambda^{(00)}_{K,K-Q}
\nonumber\\
=&&\sum_{\delta,\delta'}\sum_{Q}{\tilde A}_{\delta}(Q)
{\tilde A}_{\delta'}(-Q)e^{-i{\bf q}\cdot\delta}\sum_{K}
e^{i{\bf k}\cdot(\delta+\delta')}
\left[\Lambda^{(00)}_{K,K}+\frac{1}{2}Q^{2}\frac{\partial^{2}}{\partial Q^{2}}
\Lambda^{(00)}\Big|_{Q=0}+\cdots\right]
\nonumber\\
=&&\sum_{\delta,\delta'}\sum_{Q}{\tilde A}_{\delta}(Q)
{\tilde A}_{\delta'}(-Q)e^{-i{\bf q}\cdot\delta}\sum_{K}
e^{i{\bf k}\cdot(\delta+\delta')}\Lambda^{(00)}_{K,K}
\nonumber\\
+&&
\sum_{\delta,\delta'}\sum_{Q}{\tilde A}_{\delta}(Q)
{\tilde A}_{\delta'}(-Q)e^{-i{\bf q}\cdot\delta}\sum_{K}
e^{i{\bf k}\cdot(\delta+\delta')}
\frac{1}{2}Q^{2}\left[\frac{\partial^{2}}{\partial Q^{2}}
\Lambda^{(00)}\right]_{Q=0}+\cdots
\end{eqnarray}
where $\Lambda^{(00)}_{K,K-Q}=\mbox{tr}\left[G(k)G(K-Q)\right]$.
In the bracket $[$ $]$, we do not have odd power terms in $Q$ because
of symmetry.
The first term contributes and
gives $\theta_{i,i+\delta}\theta_{i+\delta'}$ to $S^{(2)}$. Now,
we need to see
under what condition the remaining terms are negligible. 
Let us transform back to the lattice space using
\begin{equation}
{\tilde A}_{\delta}(Q)=\int dX_{i}A_{\delta}(X_{i})e^{-iQX_{i}}
\end{equation}
Then we have, for the second term,
\begin{eqnarray}
&&\sum_{\delta,\delta'}\sum_{Q}{\tilde A}_{\delta}(Q)
{\tilde A}_{\delta'}(-Q)e^{-i{\bf q}\cdot\delta}\sum_{K}
e^{i{\bf k}\cdot(\delta+\delta')}
\frac{1}{2}Q^{2}\left[\frac{\partial^{2}}{\partial Q^{2}}
\Lambda^{(00)}\right]_{Q=0}
\nonumber\\
=&&\int dX_{i}dX'_{i}\sum_{\delta,\delta'}\sum_{Q,K}
A_{\delta}({\bf r}'_{i})A_{\delta'}({\bf r}_{i}+\delta)
e^{i{\bf k}\cdot(\delta+\delta')}
\frac{1}{2}\frac{\partial^{2}}{\partial X_{i}\partial X'_{i}}
e^{iQ(X_{i}-X'_{i})}
\left[\frac{\partial^{2}}{\partial Q^{2}}
\Lambda^{(00)}\right]_{Q=0}.
\end{eqnarray}
If the lattice size is large enough that we can change $\sum_{i}\rightarrow
\int d{\bf r}_{i}$, then assuming a periodic boundary condition
$\theta({\bf r}_{i}+L)=\theta({\bf r}_{i})$,
we have
\begin{eqnarray}
&&\int dX_{i}dX'_{i}A_{\delta}({\bf r}'_{i})A_{\delta'}({\bf r}_{i}+\delta)
\frac{\partial^{2}}{\partial X_{i}\partial X'_{i}}
e^{iQ(X_{i}-X'_{i})}
\nonumber\\
=&&\frac{\partial}{\partial X_{i}}
A_{\delta}({\bf r}_{i})A_{\delta'}({\bf r}_{i}+\delta)\Big|_{boundary}
\end{eqnarray}
In general if
\begin{equation}
\left(\frac{\partial}{\partial X_{i}}\right)^{n}
A_{\delta}({\bf r}_{i})A_{\delta'}({\bf r}_{i}+\delta)\Big|_{boundary}=0
\end{equation}
we may simply put $Q=0$ in $\Lambda^{(00)}_{K,K-Q}$ for $S^{(2)}$.
Since $\theta({\bf r}_{i})$ is periodic, so are its derivatives.
Consequently, the above condition is satisfied by the periodic
boundary condition.

\section{reduction to the continuum limit}

In this appendix we show how $\H_{\theta}$ reduces to the usual form in
the continuum limit. 
In order to do so, we need to recover the lattice
constance $a$ and take the limit $a\rightarrow0$ while keeping
$ta^{2}\rightarrow{1\over 2m}$, 
where $m$ is the electron mass, because
$\xi\rightarrow k^{2}/2m-\mu$.
Let us consider first
\bea
\H^{(1)}_{\theta}&=&
t\sum_{K}\mbox{tr}
\left[G(K){\hat\tau}_{3}e^{i\eta\omega_{n}{\hat\tau}_{3}}\right]
\cos(k_{x}a)\tau\sum_{<ij>}\left[1-\cos\left({\theta_{i,j}\over2}
\right)\right]
\nonumber\\
&\simeq&t\sum_{K}\mbox{tr}
\left[G(K){\hat\tau}_{3}e^{i\eta\omega_{n}{\hat\tau}_{3}}\right]
\left(1-{1\over2}(k_{x}a)^{2}\cdots\right)
{1\over8}\sum_{i,\delta}\left(\theta_{i}-\theta_{i+\delta}\right)^{2}
\nonumber\\
&\rightarrow&
\sum_{K}\mbox{tr}
\left[G(K){\hat\tau}_{3}e^{i\eta\omega_{n}{\hat\tau}_{3}}\right]
\int d{\bf r}{(\nabla\theta)^{2}\over 8m}\;.
\eea

Now consider $\H^{(2)}$:
\bea
\H^{(2)}&=&
{1\over2}t^{2}\sum_{\delta,\delta'}\sum_{K}\mbox{tr}\left[G(K)^{2}\right]
e^{i{\bf k}\cdot(\delta-\delta')}\sum_{i}
\sin\left({\theta_{i,i+\delta}\over2}\right)
\sin\left({\theta_{i,i+\delta'}\over2}\right)
\nonumber\\
&\simeq&
{1\over8}t^{2}\sum_{K}\mbox{tr}\left[G(K)^{2}\right]
\sum_{\delta,\delta'}\left\{
1-{a^{2}\over2}\left[{\bf k}\cdot(\delta-\delta')\right]^{2}\right\}
\sum_{i}
\left(\theta_{i}-\theta_{i+\delta}\right)
\left(\theta_{i}-\theta_{i+\delta'}\right)
\nonumber\\
&\rightarrow&
{1\over 8m^{2}}\sum_{K}\mbox{tr}\left[G(K)^{2}\right]
\int d{\bf r}\left[k^{2}_{x}(\partial_{x}\theta)^{2}+
k^{2}_{y}(\partial_{y}\theta)^{2}\right]
\nonumber\\
&=&\sum_{K}\mbox{tr}\left[G(K)^{2}\right]k^{2}
\int d{\bf r}{(\nabla\theta)^{2}\over 16m^{2}}\;.
\eea

\section{off-diagonal terms}

The effective Hamiltonian is
\bea
\H_{\theta}=&&\alpha\sum_{<ij>}\theta^{2}_{i,j}
+2\beta\sum_{i}\left(
\theta_{i,i+{\hat x}}\theta_{i,i-{\hat x}}+
\theta_{i,i+{\hat y}}\theta_{i,i-{\hat y}}\right)
\nonumber\\
+&&2\gamma\sum_{i}
\left(\theta_{i,i+{\hat x}}+\theta_{i,i-{\hat x}}\right)
\left(\theta_{i,i+{\hat y}}+\theta_{i,i-{\hat y}}\right)\;.
\eea
First let us consider 
the second term.
\bea
&&2\beta\sum_{i}\left(
\theta_{i,i+{\hat x}}\theta_{i,i-{\hat x}}+
\theta_{i,i+{\hat y}}\theta_{i,i-{\hat y}}\right)
\nonumber\\
\simeq &&2\beta\sum_{i}\left[
\sin(\theta_{i,i+{\hat x}})\sin(\theta_{i,i-{\hat x}})+
\sin(\theta_{i,i+{\hat y}})\sin(\theta_{i,i-{\hat y}})
\right]
\nonumber\\
\simeq &&2\beta\sum_{i}\left[
\left({\bf S}_{i}\times{\bf S}_{i+{\hat x}}\right)\cdot
\left({\bf S}_{i}\times{\bf S}_{i-{\hat x}}\right)+
\left({\bf S}_{i}\times{\bf S}_{i+{\hat y}}\right)\cdot
\left({\bf S}_{i}\times{\bf S}_{i-{\hat y}}\right)\right]
\eea
Applying a vector
identity to this expression, we then obtain
\be
({\bf S}_{i}\times{\bf S}_{i+{\hat x}})\cdot
({\bf S}_{i}\times{\bf S}_{i-{\hat x}})={\bf S}_{i+{\hat x}}\cdot
{\bf S}_{i-{\hat x}}-({\bf S}_{i}\cdot{\bf S}_{i+{\hat x}})
({\bf S}_{i}\cdot{\bf S}_{i-{\hat x}})\;,
\ee
where we used
the fact that ${\bf S}_{i}\cdot{\bf S}_{i}=1$,
we know that the second term
becomes
\be
\beta\sum_{<ij>_3}{\bf S}_{i}\cdot{\bf S}_{j}-
2\beta\sum_{i}\left[
\cos(\theta_{i,i+{\hat x}})\cos(\theta_{i,i-{\hat x}})+
\cos(\theta_{i,i+{\hat y}})\cos(\theta_{i,i-{\hat y}})\right]\;.
\ee
Since $\cos(\theta_{i,i+{\hat x}})\simeq 1-(\theta_{i,i+{\hat x}})^{2}/2$,
we obtain, for the second term,
\be
2\beta\sum_{i}\left(
\theta_{i,i+{\hat x}}\theta_{i,i-{\hat x}}+
\theta_{i,i+{\hat y}}\theta_{i,i-{\hat y}}\right)
\simeq\beta\sum_{<ij>}\theta^{2}_{i,j}+
\beta\sum_{<ij>_3}{\bf S}_{i}\cdot{\bf S}_{j}\;.
\ee
It is also straightforward to show that
\be
2\gamma\sum_{i}
\left(\theta_{i,i+{\hat x}}+\theta_{i,i-{\hat x}}\right)
\left(\theta_{i,i+{\hat y}}+\theta_{i,i-{\hat y}}\right)
\simeq
2\gamma\sum_{<ij>}\theta^{2}_{i,j}
+2\gamma\sum_{\ll ij\gg}{\bf S}_{i}\cdot{\bf S}_{j}\;.
\ee

\section{coupling constants}

\bea
{\cal J}_{1}&=&{1\over4}\Bigg\{
t\sum_{K}\mbox{tr}[G(K){\hat\tau}_{3}
e^{i\eta\omega_{n}{\hat\tau}_{3}}]\cos(k_{x})
+t^{2}\sum_{K}\mbox{tr}[G(K)G(K)]\Bigl[1+\cos(2k_{x})
\nonumber\\
&+&2\cos(k_{x})\cos(k_{y})
+2\left({t'\over t}\right)\cos(2k_{x})\cos(2k_{y})\Bigr]\Bigg\}\;,
\eea
where $\eta\rightarrow0^{+}$ is a convergent factor.
\bea
{\cal J}_{2}&=&{1\over4}\Bigg\{
-t'\sum_{K}\mbox{tr}[G(K){\hat\tau}_{3}]\cos(k_{x})\cos(k_{y})
\nonumber\\
&+&t^{2}\sum_{K}\mbox{tr}[G(K)G(K)]\biggl[\cos(k_{x})\cos(k_{y})\
-2\left({t'\over t}\right)\left(\cos(k_{x})+\cos(2k_{x})\cos(k_{y})\right)
\nonumber\\
&+&\left({t'\over t}\right)^{2}\left(1+\cos(2k_{x})\cos(2k_{y})
+2\cos(2k_{x})\right)\biggr]\Bigg\}
\eea
\bea
{\cal J}_{3}&=&
{1\over8}t^{2}\sum_{K}\mbox{tr}[G(K)G(K)]
\left[1+2\left({t'\over t}\right)^{2}\right]\cos(2k_{x})
\\
{\cal J}_{4}&=&
{1\over4}tt'\sum_{K}\mbox{tr}[G(K)G(K)]\cos(2k_{x})\cos(k_{y})
\eea
and
\be
{\cal J}_{5}=
{1\over8}t'^{2}\sum_{K}\mbox{tr}[G(K)G(K)]\cos(2k_{x})\cos(2k_{y})\;.
\ee
In actual calculations, we use
\be
T\sum_{\omega_{n}}
\mbox{tr}[G({\bf k},\omega_{n}){\hat\tau}_{3}
e^{i\eta\omega_{n}{\hat\tau}_{3}}]=1-{\xi_{\bf k}\over E_{\bf k}}
\tanh\left({\xi_{\bf k}\over 2T}\right)
\ee
and
\be
T\sum_{\omega_{n}}
\mbox{tr}[G({\bf k},\omega_{n})^{2}]=
-{1\over 2T}\left[1-\tanh^{2}\left({\xi_{\bf k}\over 2T}\right)
\right]
\ee
\ewt

\bibliographystyle{prb}

\begin{thebibliography}{1}
\bibitem{practically} This conception of a 2D superconductivity
is more or less for the pratical purpose. 
However,
although the superconductivity
is a 3D phenomenon, a 2D model for the \pg state is not ill-conceived.
\bibitem{gauge} This gauge transformation results in
the separation of the effective action into the mean field part and
the phase fluctuation part. Thus one can do a proper theory in 2D.
However, its drawback is that one needs to introduce a branch cut to have
a single-valued function for the phase field, which may cause some
technical complications.
However, these are beyond a scope of this paper.
\bibitem{bc} There might be a possibility
to deal with the full expansion without requiring a boundary condition.
However, we believe it is highly unlikely.
\end{thebibliography}

\end{document}